\documentclass[aps,prd,onecolumn,amssymb]{revtex4}
\usepackage{graphicx,bm,color}
\usepackage{amsmath}
\usepackage{amssymb}
\usepackage{amsfonts}
\usepackage{epsfig}
\newcommand{\be}{\begin{equation}}
\newcommand{\ee}{\end{equation}}
\newcommand{\bea}{\begin{eqnarray}}
\newcommand{\eea}{\end{eqnarray}}
\newcommand{\beaa}{\begin{eqnarray*}}
\newcommand{\eeaa}{\end{eqnarray*}}

\allowdisplaybreaks[4]

\begin{document}

\title{Dynamical non-locality in the near-horizon region of a black hole with quantum time}
\author{H. Hadi$^1$}\email{hamedhadi1388@gmail.com}
\author{Amin Rezaei Akbarieh$^1$}\email{am.rezaei@tabrizu.ac.ir}
\affiliation{ $^1$Faculty of Physics, University of Tabriz, Tabriz 51666-16471, Iran}

\begin{abstract}
	The formalization of the modular energy operator within the curved spacetime is achieved through the timeless approach proposed by Page and Wootters. The investigation is motivated by the peculiar behavior of the near horizon region of a black hole and its quantum effects, leading to a restriction of the study to the immediate vicinity. The focus lies on the perspective of a static observer positioned close to the horizon. This paper highlights the alteration of the modular energy's behavior in this region compared to flat spacetime. Furthermore, it is observed that the geometry of the spacetime influences the non-local properties of the modular energy. Moreover, within the event horizon of the black hole, the modular energy exhibits a completely distinct behavior, rendering its modular behavior imperceptible in this specific region.
\end{abstract}
\maketitle

\section{Introduction}
Non-locality has been an intriguing concept in quantum mechanics since its inception, presenting a significant challenge to the scientific community \cite{locality1,locality2}. From the beginning of quantum mechanics, researchers have extensively investigated this concept through the lens of Bell's non-locality \cite{bell1,bell2}, quantum steering \cite{steering1,steering2}, quantum entanglement \cite{entanglement1}, and quantum discord \cite{discord1,discord2,discord3,discord4}. Quantum experiments have consistently violated Bell's inequalities, providing compelling evidence that quantum mechanics fundamentally diverges from classical mechanics \cite{b12,b13,b14,b15,b16,b17,b18,b19}. 

The above-mentioned non-locality is known as kinematic non-locality, which refers to the non-locality that arises from the preparation and measurement of a system. In recent years, there have been efforts to incorporate kinematic non-locality into the study of gravity. One such proposal, known as quantum gravity induced entanglement of masses (QGEM), suggests testing the quantum nature of gravity by observing the spin entanglement between two quantum superposed test masses \cite{bose5,bose6}. Other proposals aim to test the spin-2 nature of gravitational interaction by observing entanglement between a quantum system and a photon \cite{bose8}. These experimental protocols rely on the Local Operation and Classical Communication (LOCC) theorem, which states that entanglement cannot occur between two quantum systems that were not initially entangled \cite{bose9,bose10}. If gravity follows the rules of quantum mechanics, two test masses will become entangled in the presence of gravitational interaction \cite{bose11,bose12,bose13,bose14,bose15,bose16}. In contrast, classical gravitational interaction with matter would not yield any entanglement \cite{bose11,bose12}. 

On the other hand, dynamical non-locality is based on modular variables and follows the equation of motion in a quantum system \cite{dynamic}. Despite the initial success demonstrated in the first applications to quantum information with continuous systems \cite{35,36,37,38}, these variables have not yet gathered the full attention of a substantial portion of the community \cite{39}. However, there have been attempts to investigate it in flat spacetime; for example, one can refer to  \cite{399}.

In this work, we survey the modular energy variable as dynamical non-locality in the presence of gravity. Motivated by the odd features of the background of the near-horizon region of the black hole, the modular energy operator is studied in this background. Some of the odd features of the near horizon region and time evolution of the near horizon region are considered in these references\cite{timedilation, hadi1,hadi2,bosso1,bosso2}. One peculiar characteristic of this region involves the violation of the equivalence principle in general relativity. At the same time, another distinct feature pertains to the frozen state of the vacuum within this area. Consequently, the creation of particles is impeded within this region. \cite{bosso2}.   Therefore, the modular energy variable in the vicinity of the near horizon background is examined through the Page-Wootters approach (PaW) \cite{paw1,paw2}, considering the relevant factors. This method employs a timeless perspective and deduces the Schr\"{o}dinger equation for the subsystem of interest.   

Since the utilization of the PaW mechanism is being employed, it is imperative to acknowledge that there exist specific criticisms regarding this approach  \cite{kuchar}. Within the PaW mechanism, the conditional probability exhibits several issues: (I) it results in inaccurate localization probabilities when considering a relativistic particle, (II) it violates the Hamiltonian constraints, and (III) it yields incorrect transition probabilities. The concerns raised in cases (II) and (III) have been thoroughly examined and resolved in the work by H\"{o}hn \cite{hohn}. On the other hand, the objection presented in case (I) has been addressed and resolved in  \cite{hohn2}. For a comprehensive understanding of the remaining objections mentioned in Kuchar's work \cite{kuchar}, one may refer to H\"{o}hn's publication \cite{hohn3}.

In this paper, a brief review of modular variables and non-locality is presented in section two. Subsequently, the model proposed in \cite{timedilation} is employed to investigate the modular energy variable in the context of curved space-time. Consequently, section three focuses on studying the PaW mechanism utilizing tortoise coordinates in the near horizon region of a black hole. Section four then delves into considering non-locality in time and modular energy within the near horizon region. Finally, a conclusion section is provided to summarize the findings.

\section{Modular variables and non-locality}

This section provides an overview of various modular variables and their associated properties. Among the modular variables discussed in the literature, modular position and momentum are the most prevalent \cite{35,36,37,38,39,40,41,42,43,44,45,46,47,48}. Modular position and modular momentum can be defined as follows, respectively.

\begin{equation}
   e^{i\frac{Xp_{0}}{\hslash}} 
\end{equation}
and

\begin{equation}
   e^{i\frac{Pl}{\hslash}} 
\end{equation}
where $l$ and $p_{0}$ are parameters with dimensions of length and momentum, respectively. These modular variables can be written as
\begin{equation}
    X_{mod} \equiv X ~~~ mod~2\pi \hslash I/p_{0}
\end{equation}
and 
\begin{equation}
    P_{mod} \equiv P ~~~ mod~2\pi \hslash I/l
\end{equation}
The given expression involves the identity operator denoted by \$I\$ and the position and momentum operators represented by $X$ and $P$, respectively. The examination of modular variables has been a subject of interest in various projects, particularly those related to interference effects\cite{41,mod47}.  
Modular energy, which exhibits dynamical non-locality in time, possesses distinct characteristics compared to other modular variables. While the modular position is typically associated with modular momentum, the treatment of modular energy in quantum mechanics does not involve the time operator. However, our understanding of general relativity states that energy and momentum represent a unified property of an object in spacetime, as described by the stress-energy tensor. Therefore, one can emphasize the concept of modular energy as follows
\begin{equation}
    e^{iH_{S}\tau/\hslash}
\end{equation}
where $H_{S}$ is the Hamiltonian of the system, and $\tau$ is a parameter in the units of time.  It is called a modular operator since the time parameter $\tau$ defines different modular energies and is associated with the energy mod $2\pi \hslash/\tau$.  The time derivative of modular energy may depend on temporally remote events (non-locality). To show this non-local manner, let $H=\frac{P^{2}}{2m}+V(X)$ be the Hamiltonian of a certain system. Here $m$ and $V(X)$ are the mass and potential, respectively.  Heisenberg's equations of motion lead to
\begin{equation}\label{quantumlaw}
 \frac{d}{dt}e^{iPl/\hslash}=-\frac{i}{\hslash}[e^{iPl/\hslash}, H]=-\frac{i}{\hslash}[V(X+lI)-V(X)]e^{iPl/\hslash}.  
\end{equation}
The equation above clearly shows the relationship between modular momentum and spatial distance $l$. The important point is that the information about the distance $l$ depends on the modular momentum concerning $p_{0}=2\pi \hslash /l$ for $P_{mod}=P~mod~P_{0}$. This concept may assume a significant role in phenomena characterized by periodic properties, such as time crystals \cite{mod49,mod50,mod51,mod52,mod53,mod54,mod55,mod56,mod57,mod58} or may offer insight on conservation laws in quantum mechanics\cite{mod59}. In contrast to its classical counterpart, the quantum behavior of modular momentum dictates that its time evolution is not dependent on local derivatives with respect to spatial variables. The classical counterpart of modular momentum can be defined as follows
\begin{equation}\label{classiclaw}
    \frac{d}{dt}e^{2\pi ip/p_{0}}=\{e^{2\pi ip/p_{0}},H\}=-i\frac{2\pi}{p_{0}}\frac{dV}{dt}e^{2\pi ip/p_{0}}.
\end{equation}
where $H$ is the Hamiltonian of the system of interest and $V$ is the potential. In section (4), the timeless formalism will be utilized to apply the quantum modular Hamiltonian for the near horizon region of a black hole.

\section{PaW mechanism by tortoise coordinate in the near-horizon region of a black hole}

To achieve our objective, we examine the region near the horizon of a black hole \footnote{The black hole under consideration in this context is the Schwarzschild black hole, specifically its near horizon region. Thus, when the term "black hole" is employed, it pertains to the Schwarzschild black hole. Nevertheless, it is crucial to note that, in the vicinity of a massive black hole, a tiny particle can be treated as conformally flat within the near horizon region. Consequently, the findings presented in this research can also be extended to encompass such scenarios.} from the perspective of a static observer. This observer, referred to as the Rindler observer, experiences an acceleration due to the force applied to remain stationary near the black hole's horizon. The PaW mechanism, a timeless approach, is of particular interest to us as it allows for the investigation of a subsystem of claim from the viewpoint of an observer. In this section, we explore this mechanism further, focusing on the metric of the spacetime as defined by the Rindler coordinates.
 
 \begin{equation}\label{Rindler}
 	ds^{2}=-\rho^{2}d\tau^{2}+d\rho^{2},
 \end{equation}
 where $\rho$ is the proper distance from the horizon of the black hole concerning a Rindler observer standing near the horizon region, we should note that the non-radial coordinates are disregarded, and our focus is solely on the radial motion.

The transformation of the metric $(\ref{Rindler})$ into a conformal form is undertaken due to its significant utility and convenience in considering field theory and semi-classical calculation within the vicinity of the horizon region. In doing so, the element in the conformal form is written as
 \begin{equation}\label{metrictortoise}
 	ds^{2}=e^{2z}(-d\tau^{2}+dz^{2}),
 \end{equation}
where the variation $\ln \rho=z$ is used. This conformal form of the metric $(\ref{metrictortoise})$  is called tortoise coordinate.

To achieve our objective, as elucidated in section (4), we present a comprehensive depiction of the primary image employed to examine the non-locality within a curved background, explicitly focusing on the near horizon region of the black hole. The PaW mechanism is implemented in the following manner for a total of $N$ relativistic particles within this curved spacetime. The specific intricacies of this mechanism in the near horizon region can be found in the reference \cite{timedilation}. Nevertheless, the primary illustration of the approach above is provided herein to serve our intended purpose.

A Hamiltonian constraint description of relativistic particles with internal degrees of freedom is presented in this study. The system under consideration consists of$N$ free relativistic particles possessing a set of internal degrees of freedom, denoted by the configuration variables $p_{q_{n}}$ for $(n=1,2,3,..., N)$. The particles move in curved spacetime with metric $g_{\mu\nu}$, which in this case is represented by the metric $(\ref{metrictortoise})$. The Lagrangian for the $nth$ particle and the action for the system are provided as follows,
 \begin{equation}
     S=\sum_{n}\int d\tau_{n}L_{n}(\tau_{n}),
 \end{equation}
 \begin{equation}\label{L}
   L_{n}(\tau_{n})=-m_{n}c^{2}+p_{q_{n}}\frac{dq_{n}}{d\tau_{n}}-H_{n}^{clock},  
 \end{equation} 
 where $m_{n}$ and $\tau_{n}$ represent the particle's rest mass and proper time, respectively. The particle's internal degrees of freedom are governed by the Hamiltonian $H_{n}^{clock}(q_{n},p_{q_{n}})$. These internal degrees of freedom serve as a clock, monitoring the proper time of the $nth$ particle.
 
The $nth$ particle's center-of-mass spacetime position for an internal observer is $x_{n}^{\mu}$. The differential proper time, which is parameterized about an arbitrary parameter $t_{n}$, can be determined as follows 
\begin{equation}
     d\tau_{n}=\sqrt{-g_{\mu\nu}\dot{x}_{n}^{\mu}\dot{x}_{n}^{\nu}}dt_{n},
 \end{equation}
where the over dot denotes the derivative with respect to $t_{n}$ and the constant  $c$ is equal to $1$ in this context.  Therefore, one can define the action as
\begin{equation}
    S=\sum_{n}\int dt_{n}\sqrt{-g_{\mu\nu}\dot{x}_{n}^{\mu}\dot{x}_{n}^{\nu}}L_{n}(t_{n}),
\end{equation}
which is invariant under the changes of worldline parameter $t_{n}$, since there is one-to-one correspondence between $t_{n}$ and $\tau_{n}$.  The Lagrangian $L(t)$ which is defined in the action $S=\int dt L(t)$ is obtained by
\begin{equation}\label{LL}
    L(t):=\sum_{n}\sqrt{-g_{\mu\nu}\dot{x}_{n}^{\mu}\dot{x}_{n}^{\nu}}(-m_{n}c^{2}+p_{q_{n}}\frac{dq_{n}}{d\tau_{n}}-H_{n}^{clock}).
\end{equation}
The Hamiltonian, obtained through the Legendre transform, is found to be zero as a constraint, which is derived from the Legendre transform with respect to the Lagrangian $(\ref{LL})$ and is expressed as
\begin{equation}
    H:=\sum_{n}[g_{\mu\nu}p_{n}^{\mu}\dot{x}_{n}^{\nu}+p_{q_{n}}\dot{q}_{n}]-L(t)=\sum_{n}H_{n},
\end{equation}
where $p_{n}^{\mu}:=g^{\mu\nu}\frac{\partial L(t)}{\partial \dot{x}_{n}^{\nu}}$. Finally these considerations lead to $H_{n}\approx 0$ where " $\approx$" means $H_{n}$ vanishes as constraint \cite{con}. Therefore, the $N$ constraints of  $H_{n}\approx 0$ can be given by \cite{timedilation}
\begin{equation}
    C_{H_{n}}:=g_{\mu\nu}p_{n}^{\mu}p_{n}^{\nu}+M_{n}^{2}\approx 0,
\end{equation}
where the mass function $M_{n}=m_{n}+H_{n}^{clock}$. Similar to previous studies \cite{ch1,timedilation}, the constraints mentioned can also be factorized as $C_{H_{n}}=C_{n}^{+}C_{n}^{-}$, where   
\begin{equation}\label{c3}
    C_{n}^{\pm}:=\sqrt{g_{00}}(p_{n})_{0}\pm h_{n},
\end{equation}
and  
\begin{equation}\label{hn}
    h_{n}=\sqrt{g_{ij}p_{n}^{i}p_{n}^{j}+M_{n}^{2}}.
\end{equation}

In our current scenario, the metric $(\ref{metrictortoise})$ is utilized to describe the near horizon region in $(\ref{c3})$. This forces us to employ the PaW mechanism to analyze \$N\$ relativistic particles in the curved spacetime where the physical normalized $N$ relativistic particles are defined by $|\Psi\rangle\rangle$.  By enlarging Hilbert space $\mathcal{H}_{R}$ into clock and system, we have  
\begin{equation}
	\mathcal{H}=\mathcal{H}_{A}\otimes\mathcal{H}_{R}.
\end{equation} 
The Hilbert space of subsystem of interest  consists of center-of-mass $\mathcal{H}_{n}^{cm}$ and internal degrees of freedom $\mathcal{H}_{n}^{clock}$ as follows
\begin{equation}
\mathcal{H}_{R}=\otimes_{n}\mathcal{H}_{n}^{cm}\otimes \mathcal{H}_{n}^{clock}.
\end{equation}
The total Hamiltonian of the whole system can be chosen as $C_{n}^{+}$, which is a constraint and leads to Wheel-DeWitt equation $C_{n}^{+}|\Psi\rangle\rangle=0$. In our current scenario, $C_{n}^{+}=H$, indicating the total Hamiltonian of the system, which in the PaW approach can be written as,   
\begin{equation}\label{hcrossh}
H=H_{A}\otimes I_{R}+I_{A}\otimes H_{R},
\end{equation}
where $H_{R}$ is given by  expanding the equation $(\ref{hn})$ by using the metric $(\ref{metrictortoise})$. For details of calculations, refer to \cite{timedilation}.  Therefore, the Wheeler-DeWitt equation can be represented as
\begin{equation}\label{wheeler}
H|\Psi\rangle\rangle=0,
\end{equation}
where the operator $H$ acts on the state vector $|\Psi\rangle\rangle$ leading to following  Schr\"{o}dinger equation for the system of under consideration
\begin{equation}\label{schhnz}
i \frac{d}{dt_{A}}|\psi_{s}(t)\rangle=
\sum_{n}(H_{zn}^{0}
+ H_{zn}^{cm}+ H_{zn}^{clock}
-\frac{e^{z}}{m_{n}c^{2}}[H_{zn}^{cm}\otimes H_{zn}^{clock} + \frac{1}{2}(H_{zn}^{cm})^{2}])\otimes I_{S_{\bar{n}}}|\psi_{R}(t)\rangle.
\end{equation}
The $t$ on the left-hand side of equation $(\ref{schhnz})$ was conveniently labeled as  $t_{A}$ for ease of reference and to facilitate our notation in the subsequent section. Furthermore, the index $R$ represents the rest of the universe. It should be noted that the relations utilized in equation $(\ref{schhnz})$ as follows 
\begin{eqnarray}\label{znandn}
 H_{zn}^{0}=m_{n}c^{2}e^{-z}=H_{n}^{0}e^{-z}\\\nonumber
 H_{zn}^{cm}=\frac{H^{cm}_{n}}{e^{-z}}\\\nonumber
 H_{zn}^{clock}=H_{n}^{clock}e^{-z},
\end{eqnarray}
without the index $z$, the Hamiltonian is represented in the flat Minkowski spacetime.
\section{Non-locality in time via timeless framework in the near-horizon region of the black hole }
This section is dedicated to investigating the non-local behavior of the modular energy operator in the near horizon region using the PaW approach. The importance of dynamical non-locality, specifically the modular energy operator, in quantum mechanics was emphasized in the introduction section. Exploring its application in curved spacetime could potentially offer valuable insights into the concepts of non-locality in gravity. Hence, driven by this assertion and recognizing the significant role of Kinematic non-locality concepts in both quantum mechanics and gravity, our interest lies in extending dynamical non-locality, such as the modular energy operator, to curved spacetime.

The preceding section noted that the framework endeavors to account for the temporal changes occurring within the subsystem of interest, even when the entire system remains static. The clock system, as described in \cite{time1}, uses a vector in Hilbert space $\mathcal{H}_{A}$ to represent its state. The evolution of the remaining system, which is represented by Hilbert space $\mathcal{H}_{R}$, is examined through the clock system. The entire system's state, denoted as $|\Psi \rangle\rangle$, is assumed to be closed and controlled by the Wheeler-DeWitt equation$(\ref{wheeler})$, with a Hilbert space of  $\mathcal{H}_{A}\otimes \mathcal{H}_{R}$ and is given by
\begin{equation}
(H_{A}\otimes I_{R}+I_{A}\otimes H_{R})|\Psi\rangle\rangle =0.
\end{equation}

Applying a scalar product by an eigenstate $|t_{A}\rangle$ of operator $T_{A}$ to the left of expression ($|\psi(t_{A})\rangle=\langle t_{A}|\Psi \rangle\rangle$) results in the following equation:
\begin{equation}
   i \frac{\partial}{\partial t_{A}}|\psi(t_{A})\rangle=H_{R}|\psi(t_{A})\rangle,
\end{equation}
which is the Schr\"{o}dinger equation for the subsystem being considered.

We will examine how the system with Hilbert space $\mathcal{H}_{R}$ evolves over time from the perspective of an observer who is stationary near the black hole horizon and has access to Hilbert space  $\mathcal{H}_{A}$. We will then derive the $t_{A}$-evolution of the modular energy  $e^{iH_{zn}^{clock}\tau}$  of the clock ("clock") \footnote{It should be noted that there are two distinct clocks present in the system. The first clock, denoted by $A$, is associated with the Hamiltonian $H_{A}$ and operates within the Hilbert space $\mathcal{H}_{A}$. The second clock, labeled simply as "clock", is governed by the Hamiltonian $H_{zn}^{clock}$ and is a subsystem within the Hilbert space $\mathcal{H}_{R}$, belonging to the Hilbert space $\mathcal{H}_{zn}^{clock}$. } as seen by observer A. In this analysis, we take into account the following scenario. In the previous section, we discussed Page-Wootter's mechanism for the near-horizon region of the black hole. In this mechanism, we denoted the Hilbert space as $\mathcal{H}=\mathcal{H}_{A}\otimes\mathcal{H}_{S}$. Now, let us consider a clock within the Hilbert space  $\mathcal{H}_{A}$, which can either be a physical clock or a mathematical Hilbert space with a clock state  $|t_{A}\rangle$. This clock is associated with a time observable defined as a positive operator-valued measure (POVM) as follows
\begin{equation}
    T_{A}:=\{ E_{A}(t_{A}) ~~\forall t \in G| I_{A}=\int_{G} dt_{A}E_{A}(t_{A}) \},
\end{equation}

where $E_{A}(t_{A})$ is a positive operator in $\mathcal{H}_{A}$  and $G$ is the group generated by $H_{A}$ .  The measurement of the clock $A$, resulting in the time $t_{A}$, is denoted by the clock state $|t_{A}\rangle$. In the remaining portion of the Hilbert space, denoted as $\mathcal{H}_{R}$ in our model, there exists another clock represented by the Hamiltonian$H_{zn}^{clock}$, a system with Hamiltonian$H_{zn}^{cm} + H_{zn}^{0}$, and an interaction term indicating the interaction between the system and the "clock". It is important to note that we refer to the clock in the rest of the universe as a "clock". This illustrates that the external clock $A$ represents the interaction between the rest of the universe R and its external environment if such an interaction exists. The rest of the universe R itself consists of an internal clock ("clock") and a main system equipped with the Hamiltonian$H_{zn}^{cm} + H_{zn}^{0}$.

Now, in accordance with equation $(\ref{schhnz})$ we can define  $H_{eff} \in \mathcal{H}_{R} $  as follows:
\begin{equation}
	H_{eff}=H_{zn}^{0}
	+ H_{zn}^{cm}+ H_{zn}^{clock}
	-\frac{e^{z}}{m_{n}c^{2}}[H_{zn}^{cm}\otimes H_{zn}^{clock} + \frac{1}{2}(H_{zn}^{cm})^{2}].
\end{equation}
 In our model, the $H_{eff}$ is exactly equal to $H_{R}$. This can be expressed as: 
\begin{equation}\label{total}
    H_{R}=H_{eff}=H_{zn}^{0}
	+ H_{zn}^{cm}+ H_{zn}^{clock}
	 +H^{int}_{zn}
\end{equation}
By substituting this expression into equation $(\ref{wheeler})$ and taking the scalar product with an eigenstate  $|t_{A}\rangle$ of time operator  $T_{A}$ on the left ($|\psi(t_{A})\rangle=\langle t_{A}|\Psi \rangle\rangle$) , we can obtain the following equation::
\begin{equation}
   i \frac{\partial}{\partial t_{A}}|\psi(t_{A})\rangle=H_{eff}|\psi(t_{A})\rangle. 
\end{equation}
This equation is equivalent to equation  $(\ref{schhnz})$ which defines the Schr\textbackslash{}"\{o\}dinger equation for the system $R$ with time defined by an external clock $A$.  Here, $|\psi(t_{A})\rangle$ is the usual state in quantum mechanics and can be normalized at each instant of time.  We define $T_{A}$ to be a time operator associated with clock $A$ and is given by the following commutator relation
\begin{equation}
    [T_{A},H_{A}]=i ~~~~~~and ~~~~~ H_{A}=-i \frac{\partial}{\partial t_{A}}.
\end{equation}
Therefore the $|\Psi \rangle\rangle$ can be written as
\begin{equation}
    |\Psi \rangle\rangle =\int dt_{A}|t_{A}\rangle \otimes |\psi(t_{A})\rangle. 
\end{equation}
The state $|\Psi \rangle\rangle$ is called the history state, since at every time $t_{A}$ it has information about $|\psi(t_{A})\rangle$. 

The rest of the universe  Hamiltonian $R$ $(\ref{total})$, contains  $H_{int}=H_{zn}^{int}(T_{A})$ representing the interaction between Hamiltonians $ H_{zn}^{cm}$ and $ H_{zn}^{clock}$ belonging to the rest of system $R$,  is given by
\begin{equation}\label{interaction}
	H^{int}_{zn}=-\frac{e^{z}}{m_{n}c^{2}}[H_{zn}^{cm}\otimes H_{zn}^{clock} + \frac{1}{2}(H_{zn}^{cm})^{2}].
\end{equation}

One can explicitly obtain the $t_{A}$-evolution of the modular energy  $e^{iH_{zn}^{clock}\tau}$ of clock ( "clock") from the perspective of clock $A$ using Heisenberg's equation of motion. It is important to note that $\tau$ is a parameter with units of time. Therefore we have 
\begin{eqnarray}
	\frac{d}{dt_{A}}e^{iH_{zn}^{clock}\tau}&=&-i[e^{iH_{zn}^{clock}\tau},H_{eff}(T_{clock})]\\\nonumber
 &=& -i(e^{iH_{zn}^{clock}\tau}(H_{zn}^{0}
	+ H_{zn}^{cm}+H^{int}_{zn} )-(H_{zn}^{0}
	+ H_{zn}^{cm}+H^{int}_{zn})e^{iH_{zn}^{clock}\tau}),
\end{eqnarray}
which led to the evolution of modular energy
\begin{eqnarray}\label{modular}
\frac{d}{dt_{A}}e^{iH_{zn}^{clock}\tau}=-i[H_{zn}^{0}(T_{clock}+I\tau)
+ H_{zn}^{cm}(T_{clock}+I\tau)+H^{int}_{zn}(T_{clock}+I\tau)\\\nonumber
-H_{zn}^{0}(T_{clock})
-H_{zn}^{cm}(T_{clock})-H^{int}_{zn}(T_{clock})]e^{iH_{zn}^{clock}\tau}.
\end{eqnarray}

The modular Hamiltonian's quantum dynamical equation exhibits non-local behavior and is independent of local derivatives. The function of $T_{clock}$ is reflected in  $H_{eff}$, while $e^{iH_{zn}^{clock}\tau}$ prompts translation in the time variable of "clock".  By manipulating the equation $(\ref{modular})$ we have 
\begin{eqnarray}\label{mod}
  \frac{d}{dt_{A}}e^{iH_{zn}^{clock}\tau}=&-i[H_{zn}^{0}(T_{clock}+I\tau)-H_{zn}^{0}(T_{clock})]e^{iH_{zn}^{clock}\tau}\\\nonumber
  &-i[ H_{zn}^{cm}(T_{clock}+I\tau)-H_{zn}^{cm}(T_{clock})]e^{iH_{zn}^{clock}\tau}\\\nonumber
  &-i[H^{int}_{zn}(T_{clock}+I\tau)-H^{int}_{zn}(T_{clock})]e^{iH_{zn}^{clock}\tau}.
\end{eqnarray}
According to equation $(\ref{mod})$ and  $(\ref{znandn})$ , one can reach to the following results

 \begin{itemize}
     \item $e^{iH_{zn}^{clock}}$ generates translations in the time variable of the clock ("clock"), and $H_{eff}$, in general, is a function of $T_{clock}$ 
     \item  Although it seems to violate the signaling backward in time, the uncertainty principle does not allow such a phenomenon.   
     \item          The modular energy of clock "clock" evolves according to the effective Hamiltonian $H_{eff}$ at two distinct time points, as observed from clock A. 
     \item  The quantifiable manifestation of curved spacetime geometry and its background effects can be observed in the time evolution of the modular energy operator, as viewed from the perspective of clock $A$, within the Heisenberg equation of motion.

 \end{itemize}

We will now discuss the non-locality of the modular Hamiltonian in a black hole background. As we approach the horizon$(z\rightarrow -\infty)$, the right-hand side of equation $(\ref{mod})$ becomes infinite due to the term $H_{zn}^{0}(T_{clock}+I\tau)-H_{zn}^{0}(T_{clock})$ in accordance with equation $(\ref{znandn})$. From the perspective of a clock $A$ located near the horizon region, there is a strong and high energy experienced right at the black hole's horizon. The Hamiltonian $H_{zn}^{0}$ in the term  $H_{zn}^{0}(T_{clock}+I\tau)-H_{zn}^{0}(T_{clock})$ is responsible for this high energy experience as observed by a static observer near the horizon region. Consequently, the near horizon region of a black hole exhibits non-local behavior in the modular Hamiltonian, similar to quantum mechanics, although gravity and the curved spacetime geometry also play a role in the modular Hamiltonian.

Finally, we should determine that when the right-hand side of equation $(\ref{mod})$ is dominated by infinity, the modular behavior of the Hamiltonian cannot be discerned. This implies that the translations in the time variable of the clock ("clock") do not have a significant impact, as one is only at the horizon$(z\rightarrow -\infty)$  when the first term of the right-hand side of the equation $(\ref{mod})$ diverges. Consequently, the modular behavior is not perceptible at the horizon.

\section{Conclusion}

We examined the modular characteristics of the modular Hamiltonian within the vicinity of a black hole's near-horizon region, as observed by a stationary observer positioned there. Our investigation was confined to the timeless framework proposed by Page and Wootters (referred to as the PaW mechanism), which considers modular energy. The non-local behavior of the modular energy in this region is altered, indicating that the gravitational and geometric properties of spacetime influence its non-local behavior. This alteration becomes increasingly significant as we approach the horizon, where the modular behavior and non-local nature of the modular energy become indiscernible. This indiscernibility arises because, from the perspective of a stationary observer situated in the near horizon region, the energy on the horizon becomes infinitely large. Consequently, the observer in this region cannot differentiate the dependence of the modular energy's evolution on two distinct time frames on the black hole's horizon.

\section*{Acknowledgement}

This research is supported by the research grant of the University of Tabriz (sad/779).

\section*{Data availability statement}
No new data were created or analysed in this study.



\begin{thebibliography}{99}
\bibitem{time1} D.~N.~Page and W.~K.~Wootters,
Phys. Rev. D \textbf{27}, 2885 (1983)
doi:10.1103/PhysRevD.27.2885
\bibitem{locality1} Z.~Tu, D.~E.~Kharzeev and T.~Ullrich,
Phys. Rev. Lett. \textbf{124}, no.6, 062001 (2020)
doi:10.1103/PhysRevLett.124.062001
[arXiv:1904.11974 [hep-ph]].
\bibitem{locality2}E. Schr\"{o}dinger, "Discussion of probability relations between separated systems,''
Math. Proc. Camb. Philos. Soc. 31, 555 (1935) 
\bibitem{bell1}J.~S.~Bell,
Physics Physique Fizika \textbf{1}, 195-200 (1964)
doi:10.1103/PhysicsPhysiqueFizika.1.195
\bibitem{bell2} N. Brunner, D. Cavalcanti, S. Pironio, V. Scarani, and S. Wehner, Bell nonlocality, Rev. Mod. Phys. 86, 419 (2014).
\bibitem{steering1} H. M. Wiseman, S. J. Jones, and A. C. Doherty, 
Phys. Rev. Lett. 98, 140402 (2007)
doi.org/10.1103/PhysRevLett.98.140402
\bibitem{steering2} R. Uola, A. C. S. Costa, H. C. Nguyen, and O. G\"{u}hne, 
Rev. Mod. Phys. 92, 015001 (2020)
doi.org/10.1103/RevModPhys.92.015001 
\bibitem{entanglement1}R. Horodecki, P. Horodecki, M. Horodecki, and K. Horodecki, 
Rev. Mod. Phys. 81, 865 (2009)
doi.org/10.1103/RevModPhys.81.865
\bibitem{discord1}W. H. Zurek,
Ann. Phys. (Leipzig) 9, 855 (2000)
\bibitem{discord2}H. Ollivier and W. H. Zurek,
Phys. Rev. Lett. 88, 017901 (2001)
\bibitem{discord3} L. Henderson and V. Vedral, 
J. Phys. A 34, 6899 (2001)
\bibitem{discord4} K. Modi, A. Brodutch, H. Cable, T. Paterek, and V. Vedral, 
Rev. Mod. Phys. 84, 1655 (2012) 
\bibitem{35} D. Gottesman, A. Kitaev, and J. Preskill, 
Phys. Rev. A 64, 012310 (2001).
\bibitem{36} C. Gneiting and K. Hornberger, 
Phys. Rev. Lett. 106, 210501 (2011)
\bibitem{37}P. Vernaz-Gris, A. Ketterer, A. Keller, S. Walborn, T. Coudreau, and P. Milman, 
Phys. Rev. A 89, 052311 (2014)
\bibitem{38} A. Ketterer, A. Keller, S. Walborn, T. Coudreau, and P. Milman, 
Phys. Rev. A 94, 022325 (2016)
\bibitem{39} S. Popescu, 
Nat. Phys. 6, 151 (2010)
\bibitem{399} I. L. Paiva, M. Nowakowski, and E. Cohen, 
Phys. Rev. A 105, 042207, (2022) 
\bibitem{40} S. Massar and S. Pironio, 
Phys. Rev. A 64, 062108 (2001)
\bibitem{41} Y. Aharonov and D. Rohrlich, "Quantum Paradoxes: Quantum Theory for the Perplexed"
(Wiley-VCH, Weinheim, 2005)
\bibitem{42} J. Tollaksen and Y. Aharonov, 
J. Phys.: Conf. Ser. 196, 012006 (2009)
\bibitem{43} J. Tollaksen, Y. Aharonov, A. Casher, T. Kaufherr, and S. Nussinov, 
New J. Phys. 12, 013023 (2010)
\bibitem{44} M. A. D. Carvalho, J. Ferraz, G. F. Borges, P. L. de Assis, S. Padua, and S. P. Walborn, 
Phys. Rev. A 86, 032332 (2012)
\bibitem{45} A. C. Lobo, Y. Aharonov, J. Tollaksen, E. M. Berrigan, and C. de Assis Ribeiro, 
Quantum Studies: Math. Found. 1, 97 (2014)
\bibitem{46} M. R. Barros, O. J. Farıas, A. Keller, T. Coudreau, P. Milman, and S. P. Walborn, 
Phys. Rev. A 92, 022308 (2015)
\bibitem{47} Y. Aharonov, E. Cohen, F. Colombo, T. Landsberger, I. Sabadini, D. C. Struppa, and J. Tollaksen, 
Proc. Natl. Acad. Sci. USA 114, 6480 (2017)
\bibitem{48} C. Fl\"{u}hmann, V. Negnevitsky, M. Marinelli, and J. P. Home, 
Phys. Rev. X 8, 021001 (2018) 
\bibitem{mod47}Y. Aharonov, E. Cohen, F. Colombo, T. Landsberger, I. Sabadini, D. C. Struppa, and J. Tollaksen, 
Proc. Natl. Acad. Sci. USA 114, 6480 (2017)
\bibitem{mod49}F. Wilczek, 
Phys. Rev. Lett. 109, 160401 (2012)
\bibitem{mod50} F. Wilczek, 
Phys. Rev. Lett. 111, 250402 (2013)
\bibitem{mod51} P. Bruno, 
Phys. Rev. Lett. 110, 118901 (2013)
\bibitem{mod52} P. Bruno, 
Phys. Rev. Lett. 111, 070402 (2013)
\bibitem{mod53} H. Watanabe and M. Oshikawa, 
Phys. Rev. Lett. 114, 251603 (2015)
\bibitem{mod54}V. Khemani, A. Lazarides, R. Moessner, and S. L. Sondhi, 
Phys. Rev. Lett. 116, 250401 (2016)
\bibitem{mod55}D. V. Else, B. Bauer, and C. Nayak, 
Phys. Rev. Lett. 117, 090402 (2016)
\bibitem{mod56} C. W. von Keyserlingk, V. Khemani, and S. L. Sondhi, 
Phys. Rev. B 94, 085112 (2016)
\bibitem{mod57}N. Y. Yao, A. C. Potter, I.D. Potirniche, and A. Vishwanath, 
Phys. Rev. Lett. 118, 030401 (2017)
\bibitem{mod58}J. Zhang, P. W. Hess, A. Kyprianidis, P. Becker, A. Lee, J. Smith, G. Pagano, I.-D. Potirniche, A. C. Potter, A. Vishwanath, et al., 
Nature 543, 217 (2017) 
\bibitem{mod59}Y. Aharonov, S. Popescu, and D. Rohrlich, 
Proc. Natl. Acad. Sci. USA 118, e1921529118 (2021) 
\bibitem{timedilation} H. Hadi, K. Atazadeh, F. Darabi, 
Physics Letters B 834  137471 (2022)
\bibitem{con} R. Marnelius,  
Phys. Rev. D 10, 25352553 (1974) 
\bibitem{ch1} A. Vanrietvelde, P. A. H\"{o}hn, F. Giacomini, E. CastroRuiz,  
Quantum 4, 225 (2020) 
\bibitem{b12} S. J. Freedman and J. F. Clauser, 
Phys. Rev. Lett. 28, 938 (1972)
\bibitem{b13} J. F. Clauser and A. Shimony, 
Rep. Prog. Phys. 41, 1881 (1978)
\bibitem{b14} A. Aspect, P. Grangier and G. Roger, 
Phys. Rev. Lett. 47, 460 (1981)
\bibitem{b15} A. Aspect, P. Grangier and G. Roger, 
Phys. Rev. Lett. 49, 91 (1982)
\bibitem{b16} A. Aspect, J. Dalibard, and G. Roger, 
Phys. Rev. Lett. 49, 1804 (1982)
\bibitem{b17} M. Genovese, 
Phys. Rep. 413, 319 (2005)
\bibitem{b18} L. K. Shalm, E. Meyer-Scott, B. G. Christensen, P. Bierhorst, M. A. Wayne, M. J. Stevens, T. Gerrits, S. Glancy, D. R. Hamel, M. S. Allman, et al., 
Phys. Rev. Lett. 115, 250402 (2015)
\bibitem{b19} M. Giustina, M. A. Versteegh, S. Wengerowsky, J. Handsteiner, A. Hochrainer, K. Phelan, F. Steinlechner, J. Kofler, J. A. Larsson, C. Abellan, et al., 
Phys. Rev. Lett. 115, 250401 (2015) 
\bibitem{bose5} S. Bose, A. Mazumdar, G. W. Morley, H. Ulbricht, M. Toros, M. Paternostro, A. Geraci, P. Barker, M. S. Kim, and G. Milburn, 
Phys. Rev. Lett. 119, 240401 (2017)
\bibitem{bose6} https://www.youtube.com/watch?v=0Fv-0k13s\_k
(2016), accessed 1/11/22. 
\bibitem{bose8}D. Biswas, S. Bose, A. Mazumdar, and M. Toros, (2022),arXiv:2209.09273 [gr-qc].
\bibitem{bose9} C. H. Bennett, D. P. DiVincenzo, J. A. Smolin, and W. K. Wootters, Physical Review A 54, 3824 (1996).
\bibitem{bose10} S. Hill and W. K. Wootters, Phys. Rev. Lett. 78, 5022 (1997), arXiv:quant-ph/9703041.
\bibitem{bose11} R. J. Marshman, A. Mazumdar, and S. Bose, Phys. Rev. A 101, 052110 (2020), arXiv:1907.01568 [quant-ph].
\bibitem{bose12} S. Bose, A. Mazumdar, M. Schut, and M. Toroˇs, Phys. Rev. D 105, 106028 (2022), arXiv:2201.03583 [gr-qc].
\bibitem{bose13} D. Carney, P. C. E. Stamp, and J. M. Taylor, Classical and Quantum Gravity 36, 034001 (2019).
\bibitem{bose14}A.~Belenchia, R.~M.~Wald, F.~Giacomini, E.~Castro-Ruiz, \v{C}.~Brukner and M.~Aspelmeyer,
Phys. Rev. D \textbf{98}, no.12, 126009 (2018)
doi:10.1103/PhysRevD.98.126009
[arXiv:1807.07015 [quant-ph]].
\bibitem{bose15} D. L. Danielson, G. Satishchandran, and R. M. Wald, 
Phys. Rev. D 105, 086001 (2022)
\bibitem{bose16}M. Christodoulou, A. Di Biagio, M. Aspelmeyer, v. Brukner, C. Rovelli, and R. Howl, 
Phys. Rev. Lett. 130, 100202 (2023)
\bibitem{dynamic}Y. Aharonov, H. Pendleton, and A. Petersen, 
Int. J. Theor. Phys. 2, 213 (1969) 
\bibitem{hadi1}H. Hadi, F. Darabi, 
The European Physical Journal C , 82, 6, 537 (2022)
\bibitem{hadi2}H.Hadi, F. Darabi, 
arXiv preprint arXiv:2206.10387 
\bibitem{bosso1}L. Susskind, 
(2012)
\bibitem{bosso2} R. Bousso, 
Phys. Rev. Lett. 112, 041102 (2014)
\bibitem{paw1}D.N. Page, W.K. Wootters, 
Phys. Rev. D 27, 2885 (1983)
\bibitem{paw2}W.K. Wootters, 
Int. J. Theor. Phys. 23, 701 (1984) 
\bibitem{kuchar}K. V. Kuchar, Time and interpretations of quantum gravity,
Proc. 4th Canadian Conference on General Relativity and Relativistic Astrophysics, eds. G. Kunstatter, D. Vincent, and J. Williams (World Scientific, Singapore, 1992), pg. 69-76.
\bibitem{hohn}P.A.Höhn, A.R.H.Smith and M.P.E.Lock, 
Front. in Phys. 9, 181 (2021)
doi:10.3389/fphy.2021.587083.
\bibitem{hohn2}P.A.Höhn, A.R.H.Smith and M.P.E.Lock, 
Front. in Phys. 9, 181 (2021)
doi:10.3389/fphy.2021.587083.
\bibitem{hohn3}R. Gambini, R. A. Porto, J. Pullin, and S. Torterolo , 
Phys. Rev. D 79, 041501 (2009)




  

  




  

  

\end{thebibliography}
\end{document}